\documentclass[
 aip,
 pop,
 amsmath,amssymb,
 superscriptaddress,
 reprint,%
  nofootinbib,
]{revtex4-1}

\usepackage{graphicx}
\usepackage{dcolumn}
\usepackage{bm}
\usepackage[dvipsnames]{xcolor}
\usepackage{enumerate}
\usepackage{hyperref}
\hypersetup{colorlinks=true}
\hypersetup{citecolor=blue}

\newcommand{\mvec}[1]{\mathbf{#1}}

\newcommand{\gke}{{\tt Gkeyll}}

\begin{document}

\preprint{AIP/123-QED}

\title[Continuum Kinetic and Multi-Fluid Simulations of Classical Sheaths]{
Continuum Kinetic and Multi-Fluid Simulations of Classical Sheaths}

\author{P. Cagas}
  \affiliation{Department of Aerospace and Ocean Engineering, Virginia 
    Tech, Blacksburg, VA 24060.}
\author{A. Hakim}
  \affiliation{Plasma Physics Laboratory, Princeton University, Princeton, NJ 
    08544.}
\author{J. Juno}
  \affiliation{Institute for Research in Electronics and Applied
    Physics, University of Maryland, College Park, MD 20742.}
\author{B. Srinivasan}
  \email{srinbhu@vt.edu}
  \affiliation{Department of Aerospace and Ocean Engineering, Virginia 
    Tech, Blacksburg, VA 24060.}
\date{\today}

\begin{abstract}
  The kinetic study of plasma sheaths is critical, among other things,
  to understand the deposition of heat on walls, the effect of
  sputtering, and contamination of the plasma with detrimental
  impurities. The plasma sheath also provides a boundary condition and
  can often have a significant global impact on the bulk plasma. In
  this paper, kinetic studies of classical sheaths are performed with
  the continuum kinetic code, \gke, that directly solves the
  Vlasov-Maxwell equations. The code uses a novel version of the
  finite-element discontinuous Galerkin (DG) scheme that conserves
  energy in the continuous-time limit.  The fields are computed using
  Maxwell equations. Ionization and scattering collisions are
  included, however, surface effects are neglected. The aim of this
  work is to introduce the continuum kinetic method and compare its
  results to those obtained from an already established finite-volume
  multi-fluid model also implemented in \gke. Novel boundary
  conditions on the fluids allow the sheath to form without specifying
  wall fluxes, so the fluids and fields adjust self-consistently at
  the wall. The work presented here demonstrates that the kinetic and
  fluid results are in agreement for the momentum flux, showing that
  in certain regimes, a multi-fluid model can be a useful
  approximation for simulating the plasma boundary.  There are
  differences in the electrostatic potential between the fluid and
  kinetic results.  Further, the direct solutions of the distribution
  function presented here highlight the non-Maxwellian distribution of
  electrons in the sheath, emphasizing the need for a kinetic model.
   Densities, velocities, and potential show good
    agreement between the kinetic and fluid results.  However, kinetic
    physics is highlighted through higher moments such as parallel and
    perpendicular temperatures which provide significant differences
    from the fluid results in which the temperature is assumed to be
    isotropic.  Besides decompression cooling, the heat flux is
    shown to play a role in the temperature differences that are
    observed, specially inside the collisionless sheath.
\end{abstract}

\maketitle

\section{\label{sec:introduction}Introduction}

When plasma is contained by walls, the boundaries behave as sinks. Due
to their high thermal velocity (in comparison to heavier ions)
electrons are quickly absorbed into the wall, which leads to the
creation of a typically positive space charge region called a
sheath.\cite{Robertson2013} The resulting potential barrier works to
equalize fluxes to the wall. Even though the sheath width is usually
on the order of a Debye length, $\lambda_D$, it plays an important role
in particle, momentum, energy and heat transfer, and surface
erosion, which can, in turn, have a global effect on the
plasma. Furthermore, field-accelerated ions and hot electrons are
known to cause an emission from the solid surface that can further
alter the system.  Therefore, the sheath must be self-consistently
included and resolved in numerical simulations. This significantly
affects a computational cost of simulations, because the scale length
of the system is usually several orders of magnitude higher than the
Debye length. Usually, the effect of the sheath is mimicked with
``sheath boundary conditions'', often constructed from very simple
flux balance arguments or making assumptions like cold ions and no
surface effects.\cite{Loizu2012} Hence, first-principle simulations
of the sheath are needed to both validate and further develop the
simple models as well as to understand the global kinetic effects of
sheaths on the bulk plasma.

Sheath physics has been studied since early the works of
Langmuir,\cite{Langmuir1923} but some processes remain to be fully
understood. The original criterion for a shielding sheath, commonly
known as the Bohm criterion\cite{Bohm1949} is given by
\begin{equation}\label{eq:BohmCriterion}
    u^2_{i,0} \geq \frac{\text{k}_\text{B}T_e}{m_i},
\end{equation}
where $u_{i,0}$ is the ion bulk velocity perpendicular to the wall at
the sheath edge. The Bohm criterion assumes mono-energetic ions,
Boltzmann electrons, and no sources in the plasma, and does not depend
on the ion distribution function at the edge. Effectively, it assumes
a single-component fluid with ion inertia and electron pressure. The
Bohm criterion then requires for ions to be accelerated in the
presheath to the speed of ion acoustic waves.\cite{Riemann1990}
Surprisingly, even with the assumptions mentioned above, the Bohm
criterion applies to conditions beyond these assumptions, with errors
within 20\thinspace-\thinspace30\thinspace\%.\cite{Bohm1949}

Kinetic effects (ions are no longer monoenergetic but are rather
distributed over the velocity space; electrons no longer instantly
follow the electrostatic potential) are incorporated in the
Tonks-Langmuir model with the solution presented in
Ref\thinspace[\onlinecite{Harrison1959}]. This ``kinetic Bohm
criterion'' is discussed and generalized in several
papers\cite{Allen1976, Bissell1987, Riemann1990, Riemann1995,
  Fernsler2005, Riemann2006} and its applicability on different plasma
distribution functions is further addressed.\cite{Baalrud2011,
  Riemann2012, Baalrud2012} An alternative approach based on the fluid
moment hierarchy is presented in recent work.\cite{Baalrud2014} The
latter approach leads to a sheath criterion that is similar to the
original Bohm criterion but contains an extra term for the ion
temperature.

When the energy of incident particles is high enough, bounded
electrons in the wall can be ejected, and the boundary begins acting
as a source of plasma. This mechanism, known as secondary electron
emission (SEE), is critical for devices like Hall
thrusters\cite{Dunaevsky2003} and tokamak walls.\cite{Takamura2004}
Particle-in-cell (PIC) simulations\cite{Sydorenko2004, Sydorenko2006}
show that the electron distribution function in Hall thrusters is, due
to the SEE, strongly anisotropic and depletes at high
energies. Therefore, a kinetic approach is required. Further
discussion of kinetic effects, plasma flux to the wall, secondary
electron fluxes, plasma potential, and electron cross-field
conductivity are presented in
Ref\thinspace[\onlinecite{Kaganovich2007}]. Recent
work\cite{Campanell2012} studies conditions for sheath instability due
to SEE and ``weakly confined electrons'' at the boundary of the loss
cone.

PIC simulations and Direct Simulation Monte-Carlo (DSMC) are currently
the most widely used methods for kinetic scales. They are robust,
allow complex geometries, and can include a broad range of physical
and chemical processes. However, PIC simulations are subject to noise
-- an issue that can be overcome by using continuum kinetic solvers
with high-order accurate algorithms. Also, simulations of complex
problems like Hall thrusters have relevant scales from the
magneto-hydrodynamic (MHD) scale to the kinetic scale and efficiently
resolving all of them is computationally expensive.  Continuum kinetic
solvers potentially allow easier implementation of hybrid
(fluid-kinetic) algorithms for problems requiring such a scale
separation, and enable the use of novel multi-scale techniques like
asymptotic preserving methods\cite{Emako2016,Liu2016} that
self-consistently transition between the regimes without changing the
underlying equations or implementation.

Here the focus is on directly solving the Boltzmann equation,
\begin{equation}\label{eq:Boltzmann}
  \frac{\partial f}{\partial t} + \textbf{v}\cdot\frac{\partial
    f}{\partial\textbf{x}} + \frac{q}{m}\left(\textbf{E} +
    \textbf{v}\times\textbf{B}\right)\frac{\partial
      f}{\partial\textbf{v}} = S,
\end{equation}
using a discontinuous Galerkin (DG) scheme.\cite{Cockburn2001} Here,
$f(\mvec{x},\mvec{v},t)$ is the particle distribution function, $q$ is
charge, $m$ is mass and $S$ represents different source terms. A
continuum Eulerian scheme is used in this work. DG schemes have been
extensively developed and used in the computational fluid dynamics and
applied mathematics community over the past 15 years, as they combine
some of the advantages of finite-element schemes (low phase error,
high accuracy, flexible geometries) with finite-volume schemes
(limiters to preserve positivity/monotonicity, locality of computation
for parallelization).

The kinetic DG results are compared to results from a two-fluid
finite-volume scheme.\cite{Hakim2006} The two-fluid model uses novel
boundary conditions that compute fluxes self-consistently using
Riemann solvers at the wall. Therefore, the fluxes to the wall are a
result of the model rather than input parameters. This work
demonstrates an agreement between kinetic and fluid solutions in
density and momentum for several regimes. The sheath potential differs
on order of 10\%, and the difference is more pronounced as ion
temperature becomes significant. More significant
  differences are observed in the temperature profiles.  The kinetic
  results, performed using one configuration space dimension and two
  velocity space dimensions, produce parallel ($x$-direction) and
  perpendicular electron temperatures that differ significantly from
  the isotropic fluid temperature. This difference is attributed to
  the parallel heat flux, not present in the five-moment two-fluid
  model.  Furthermore, the smooth distribution function profiles
obtained from the kinetic model highlight the non-Maxwellian nature of
the species distribution inside the sheath. This is a fundamental
difference between kinetic and fluid sheath simulations as fluid
models assume a Maxwellian distribution throughout.

This paper is organized as follows. Section\thinspace\ref{sec:sheath}
provides a short review of sheath theory and the Bohm
criterion. Section\thinspace\ref{sec:kinetics} covers the continuum
kinetic model and section\thinspace\ref{sec:fluid} briefly describes
the electromagnetic five-moment two-fluid model.  Benchmarking of the
code with simulations of the Weibel instability is presented in
section\thinspace\ref{sec:bench}.  Problem setup and initial
conditions are discussed in section\thinspace\ref{sec:setup}.
Comparison between kinetic and fluid models is provided in
section\thinspace\ref{sec:comparison} and finally
section\thinspace\ref{sec:conclusions} summarizes this work and
outlines future plans.

\section{\label{sec:sheath}Brief review of sheath theory}

Langmuir's two-scale description\cite{Langmuir1923} is used here,
which assumes a quasi-neutral presheath and a sheath region with
non-zero space charge. This description is valid as long as
$\lambda_D/L \ll 1$. The dimensionless parameters in this section are
consistent with previous publications,\cite{Riemann1990}
\begin{align*}\label{eq:normunits}
    y &= \frac{m_i v_i^2}{2\text{k}_\text{B}T_e}, & \chi &=
    -\frac{e\phi}{\text{k}_\text{B}T_e}, \\ 
    n_{e,i} &= \frac{N_{e,i}}{N_0}, & \xi &= \frac{x}{\lambda_D},
\end{align*}
where $N_{e,i}$ are electron and ion number densities and $N_0$ is
charged particle number density at the sheath edge (same for electrons
and ions in the $\lambda_D/L \rightarrow 0$ approximation). The
simulations use a slightly modified set of variables that are more
suitable for practical implementation. The classical Bohm criterion
($y_0 \geq \frac{1}{2}$) can then be derived from the ion continuity
equation, $n_iy^{1/2} = y_0^{1/2}$, ion energy conservation
$y = y_0+\chi$, Boltzmann electron density $n_e = \text{exp}(-\chi)$,
and the Poisson equation
$\text{d}^2\chi/\text{d}\xi^2 = n_i-n_e$.\cite{Riemann1990}

Since the Bohm criterion applies a boundary condition on the ion drift
speed, there is a need for a mechanism that would accelerate ions in
the presheath. This mechanism is described as an ambipolar diffusion
with nonlinear effects caused by the ion inertia.\cite{Persson1962}

Equating the ion density with the electron Boltzmann factor,
\begin{align}
  n_i\sqrt{y} = j_i, \quad  
  j_i = \left(\frac{m_i}{2\text{k}_\text{B}T_e}\right)^{1/2}\frac{J_i}{N_0}
\end{align}
where $J_i$ is ion current density gives
\begin{equation}\label{eq:sheath:presheath}
  \frac{\text{d}y}{\text{d}\zeta}-\frac{\text{d}\chi}{\text{d}\zeta} <
  \frac{1}{j_i} \frac{\text{d}j_i}{\text{d}\zeta},
\end{equation}
for the presheath ($y_0 < \frac{1}{2}$). Here $\zeta = x/L$ where $L$
is a characteristic scale in the system.

From Eq.\thinspace(\ref{eq:sheath:presheath}) and the conservation of
energy, $y = y_0+\chi$, the acceleration to the Bohm velocity is
possible only if\cite{Riemann1990}
\begin{enumerate}
  \item $\text{d}j_i/\text{d}\zeta > 0$ and/or
  \item $\text{d}y/\text{d}\zeta < \text{d}\chi/\text{d}\zeta$.
\end{enumerate}
These conditions can be fulfilled by several mechanisms. Relevant to
this work are the collisional presheath and ionizing presheath. The
collisional presheath introduces ion friction and therefore fulfills
$\mathrm{d}y/\mathrm{d}\zeta < \mathrm{d}\chi/\mathrm{d}\zeta$ with
$L$ being ion mean free path. The ionizing presheath satisfies both
the increase in current gradient condition, $\text{d}j_i/\text{d}\zeta
> 0$, and ion retardation because the particles created by ionization
are assumed to have zero drift velocity. In this case, $L$ corresponds
to the mean ionization path. The model presented here includes both
scattering collisions and electron impact ionization.

Ref\thinspace[\onlinecite{Baalrud2014}] emphasizes that the concept of
a sheath edge in Langmuir's description is connected strictly to the
charge density and therefore should be independent of a plasma model
(an example of a description that is dependent on a plasma model is
the Child-Langmuir formula). Instead they suggest identifying the
sheath edge using a threshold for the normalized charge density
$\bar{\rho}_c = (n_i-n_e)/n_i$. However, in a real situation where
$\lambda_D/L \neq 0$ this transition is not abrupt, hence, arbitrary
values must be chosen.  In the results presented using the continuum
kinetic model, the thresholds for $\bar{\rho}_c = 1$\thinspace\% and
10\thinspace\% are marked by arrows. By taking the expansion of $\rho$
with respect to $\phi$, the quantitative form of the sheath condition
is derived,\cite{Baalrud2014}
\begin{equation}\label{eq:sheathEdge}
  \left|\frac{\mathrm{d}n_i}{\mathrm{d}x}\right| \leq
  \left|\frac{\mathrm{d}n_e}{\mathrm{d}x}\right|.
\end{equation}

The density gradients required for Eq.\thinspace(\ref{eq:sheathEdge})
can be obtained from the steady Boltzmann equation
leading to a slightly modified version of the classical Bohm
criterion\cite{Baalrud2014}
\begin{equation}
  u_i \geq v_\mathrm{B} = \sqrt{\frac{\mathrm{k}_\mathrm{B}\left(T_e +
      T_i\right) - m_ev_e^2}{m_i}}.
\end{equation}

\section{\label{sec:kinetics}Continuum kinetic model}
\subsection{Numerical method}
An energy-conserving, discontinuous Galerkin (DG) scheme is used to
discretize the Valsov-Maxwell equations. The time-derivative term is
discretized with a strong-stability preserving Runge-Kutta scheme. In
the time-continuous limit, with a specific choice of numerical flux
for Maxwell equations, the scheme can be shown to conserve total
(particle plus field) energy. Although momentum is not conserved
exactly, the errors in momentum conservation are independent of
velocity space resolution and converge rapidly with increasing
configuration space resolution. While developing the code significant
numerical advance have been made, allowing efficient solution of the
Vlasov-Maxwell (and hybrid moment-Vlasov-Maxwell) systems. These
developments will be presented in a forthcoming paper.

\subsection{\label{sec:ioncolKin}Ionization and collision terms}

Discussion in section\thinspace\ref{sec:sheath} shows that adding an
ionizing source term can provide a steady-state.\cite{Riemann1990} A
simplified version of electron impact ionization is derived from the
exact operator of the form\cite{Meier2012}
\begin{equation}
  S_{ion,s} =
  f_n(\textbf{v})\int_{-\infty}^{\infty}\sigma(|\textbf{v}-\textbf{v}'|)
   |\textbf{v}-\textbf{v}'| f_e(\textbf{v}') \mathrm{d}\textbf{v}',
\end{equation}
where $\sigma(|\textbf{v}-\textbf{v}'|)$ is the ionization
differential cross-section with units of $[L]^2$. Assuming the
electron thermal velocity is high in comparison to the relative bulk
speed, the relative speed, $|\textbf{v}-\textbf{v}'|$, can be
approximated by the electron random velocity. The formula then
simplifies to
\begin{equation}\label{eq:collision1}
  S_{ion,s} \approx f_n(\textbf{v}) \left\langle \sigma
  v_e\right\rangle n_e,
\end{equation}
where $\left\langle \sigma v_e\right\rangle$ is the value averaged
over the velocity space.

The neutral distribution is assumed to be a non-drifting bi-Maxwellian
\begin{equation}
  f_{BM,s} = \frac{n_s}{\sqrt{2\pi v_{thx,s}^2}\sqrt{2\pi
      v_{thy,s}^2}} e^\frac{-v_x^2}{2v_{thx,s}^2}
  e^\frac{-v_y^2}{2v_{thy,s}^2},
\end{equation} 
that does not get depleted during the course of the
simulation. 
Assuming that species created by ionization are thermalized on
time-scales shorter than other collision times, and the neutral number
density is not a function of time,
Eq.\thinspace(\ref{eq:collision1}) is rewritten as
\begin{equation}
    S_{ion,s}  =
    \left\langle \sigma v_r\right\rangle n_n
    f_{BM}\left(n_e,\,0,\,\textbf{v}_{th,s}\right).\label{eq:ionization}
\end{equation}
On the right-hand-side there is a distribution function with local
number density of electrons, zero drift velocity, and local temperature of
respected species. 

\begin{figure}[!htb]
  \includegraphics[width=0.9\linewidth]{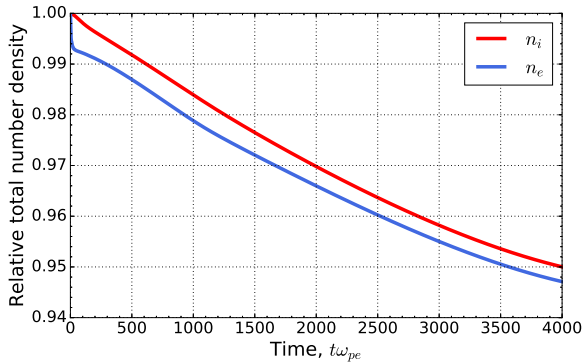}
  \caption{Relative integrated number densities of electrons and ions
    evolved over $4000/\omega_{pe}$, showing that the ionization and
    collision terms nearly balance.}
  \label{fig:totalPtcl}
\end{figure}

The ionization rate, $\left\langle \sigma v_r\right\rangle$ in
Eq.\thinspace({\ref{eq:ionization}}), is selected so as to achieve a
quasi-steady-state. Using a cold ion and Boltzmann electron model, a
simple balance of ionization sources and particle loss at the walls
shows that to achieve steady-state the ionization rate must
be\cite{Stangeby2000}
\begin{equation}\label{eq:kinetic:ionization}
  \left\langle \sigma v_r\right\rangle =
  \frac{2u_B}{L}\left(\frac{\pi}{2}-1\right),
\end{equation}
where the factor of 2 corresponds to the two-wall setup with particles
leaving the domain on both sides (see section\thinspace\ref{sec:setup}
for more details). Even though this result is derived using
significant approximations, including even a simple ionization term
helps with density conservation. Fig.\thinspace\ref{fig:totalPtcl}
presents the relative total number density as a function of time for
ions and electrons in one of our sheath simulations. After the
simulation has progressed for 4000 plasma oscillation times
($1/\omega_{pe}$), the relative difference in initial and final
integrated number densities in the simulation for both species is less
than 5\%. Even though the number of particles is not conserved
exactly, adding this physics-based source term could lead to some
results, that are otherwise unobservable (for example the positive
sheath transition described in the
Ref\thinspace[\onlinecite{Campanell2016}]). However, achieving the
true steady-state with this approach requires a calculation of the
inelastic collision integrals and inclusion of the surface physics
effects.

To balance the loss of high-energy electrons to the walls, collisions
must be included to replenish the electron tails if steady-state is to
be achieved. These collisions, however, should be infrequent enough
that the collisional mean-free-path is much longer than the sheath
width, allowing for proper simulation of collisionless sheaths. The
presented work uses a simple BGK\cite{Bhatnagar1954} operator
\begin{equation}
  S_{coll,s} = \nu_{coll}\left(f_{M,s}-f_s\right),
\end{equation}
where $f_{M,s}$ is a Maxwellian distribution function, which is
constructed using the first three moments of $f_s$. The temperature of
the distribution is taken as $(T_x + 2T_\perp)/3$. Collision frequency
$\nu_{coll}$ is calculated locally as
\begin{equation}
  \nu_{coll, ss} =
  \frac{e^4}{2\pi\epsilon_0^2m_s^2}\frac{n_s}{v_{th,s}^3}\mathrm{ln}(\Lambda).
\end{equation}
Local values of density, $n_s$, and thermal velocity, $v_{th,s}$, are
used. The Coulomb logarithm is approximated as
$\mathrm{ln}(\Lambda)=10$. Note that the mass has a power of 2, which
is caused by the fact that thermal velocities are used instead of
temperatures. Interspecies collisions are neglected in this work. With
this setup, the collision frequency decreases as plasma enters the
sheath region. Using the initial values for electrons, the
electron-electron collision frequency near the wall is around
0.3~$\omega_{pe0}$. 

\section{\label{sec:fluid}Electromagnetic five-moment two-fluid model}

The two-fluid plasma model is included for comparison with the
continuum kinetic simulation. It includes electron inertia, separate
electron and ion temperatures, and allows for non-neutral effects
(which are crucial for obtaining Debye sheaths). The fields are
computed from the full Maxwell equations, which allows for the
inclusion of displacement currents. Each species of the plasma is
described by moments of a distribution function -- number density,
momentum, and energy\cite{Hakim2006}
\begin{gather}
  \frac{\partial n_s}{\partial t} + \frac{\partial}{\partial
    x}\left(n_s u_x\right) =
  0,\label{eq:fluid:density}\\ m\frac{\partial n_s u_x}{\partial t} +
  \frac{\partial}{\partial x}\left(p_s + m_sn_su_x^2\right) =
  n_sq_eE_x,\\ \frac{\partial\mathcal{E}_s}{\partial t} +
  \frac{\partial}{\partial x}\left(u_xp_s + u_x \epsilon_s\right) =
  q_sn_su_xE_x\label{eq:fluid:energy},
\end{gather}
where $p_s$ is the isotropic pressure for each species and $E_x$ is
the electric field. To close the system, heat flow is neglected and a
scalar closure is used for the energy $\epsilon_s$
\begin{equation}
  \epsilon_s = \frac{p_s}{\gamma-1} + \frac{1}{2}n_s u_x^2,
\end{equation}
where $\gamma = \frac{5}{3}$.

To solve the system of five-moment equations, a second order, locally
implicit scheme is used. This scheme is partly described in
Ref.\thinspace[\onlinecite{Hakim2006, Loverich2013}]. Using a
locally implicit, operator splitting approach, the time-step
restriction due to the plasma frequency and Debye length scales can be
eliminated.  This decreases the computational time for multi-fluid
simulations, especially with realistic electron/ion mass ratios, even
when using an explicit scheme. For the hyperbolic homogeneous part of
the equations, a finite-volume (FV) wave-propagation scheme is
used.\cite{Leveque2002} This scheme is based on solving the
Riemann problem at each interface to compute numerical fluxes, which
are then used to construct a second-order scheme. To ensure that the
number density remains positive, on detection of a negative
density/pressure state, the homogeneous step is recomputed using a
diffusive, but positive, Lax-flux.\cite{Bouchut2004} Although Lax-flux
adds diffusion, the scheme still conserves particles and energy.

To apply boundary conditions at the walls, a vacuum is assumed just
outside the domain (i.e. just inside the wall the plasma is
neutralized and the fields vanish). This introduces a jump in the
fluids and electric field, which is then used in a Riemann solver to
compute the self-consistent fluxes corresponding to that jump. This
ensures that the solution automatically adjusts to give the correct
surface fluxes and fields, and one does not need to specify the flux
using the Bohm criteria, as is usually done in fluid codes. In codes
that use ``ghost cells'' for the boundary conditions, this BC is
particularly easy to implement: one simply sets the fluid quantities
to zero and then uses the scheme's Riemann solver to compute the
surface fluxes that are then used to update the cell adjacent to the
wall.

\section{\label{sec:bench}Benchmarking kinetic and fluid models with Weibel instability}

As the algorithms are developed, systematic benchmarking exercises are
performed. The detailed description of the algorithms and benchmarks
for the kinetic code will be published in the future. Versions of the
kinetic algorithms, applied to the quasi-neutral gyrokinetic equations
have been benchmarked.\cite{Shi2015} The fluid code has also been
thoroughly tested, as well as used in production runs for magnetic
reconnection,\cite{Wang2015,Ng2015} global magnetosphere
simulations, Rayleigh-Taylor instability, and other problems. Both the
fluid and kinetic algorithms are implemented in the \gke\ code, and
use common infrastructure for input/output, parallelization,
simulation drivers and output visualization. \gke\ also has advanced
fluid models, including the ten-moment model\cite{Hakim2008} with
a local as well as non-local closure, a gyrokinetic model, and a full
Vlasov-Maxwell solver. This opens up the possibility of performing
hybrid sheath simulations in the future, with some particle species
evolved using fluids, while others evolved with a kinetic or reduced
kinetic model.

A benchmark comparison (against analytical theory as well as between
models) for the Weibel\cite{Weibel1959} instability is presented here.
The Weibel instability is potentially relevant to sheath formation in
some regimes,\cite{Tang2011} which motivates its selection as a
benchmark here.  Furthermore, this problem highlights the unique
features of the five-moment fluid code used here to capture particular
kinetic effects by the inclusion of multiple ``streams'' of a single
species, in this case, the electrons.

The Weibel instability is driven by anisotropic pressure, an example
of which occurs in the presence of two counter-streaming electron
beams. When a small perturbation of magnetic field is introduced
perpendicular to the relative drift velocity, the Weibel instability
causes this magnetic field to grow exponentially. In the linear
regime, the dispersion relation for this instability takes the form
\begin{equation}\label{eq:weibel:dispersion}
 \frac{1}{2} = \frac{\omega_0^2}{c^2k^2} \left[\zeta
  Z(\zeta)\left(1+\frac{v_D^2}{v_{th}^2}\right)
  +\frac{v_D^2}{v_{th}^2}\right]+\frac{v_{th}^2}{c^2}\zeta^2,
\end{equation}
where $\omega_0$ is the plasma oscillation frequency, $c$ is the speed
of light, $k$ is the instability wavenumber, $v_D$ is the populations
drift speed, and $v_{th}$ is thermal speed, $\zeta =
\omega/(\sqrt{2}v_{th}k)$, where $\omega = \omega_r +
\mathrm{i}\gamma$. $Z(\zeta)$ is the plasma dispersion function
defined as
\begin{equation}
  Z(\zeta)
  = \frac{1}{\sqrt{\pi}} \int_{-\infty}^{\infty}\frac{e^{-x^2}}{x-\zeta}
   \mathrm{d}x.
\end{equation}
In the cold fluid limit, using the asymptotic expansion of $Z(\zeta)$
for large $\zeta$, the cold fluid Weibel dispersion relation is obtained
\begin{align}
  \frac{{\omega}^{4}}{2{k}^{2}} - \left( \frac{1}{2} + \frac{1}{{k}^{2}} \right)
  {\omega}^{2} -v_{D}^{2}  = 0 \label{eq:weibel:cold-dispersion}
\end{align}
where $\omega$ is normalized to $\omega_0$, $k$ is normalized to
$\omega_0/c$ and $v_D$ is normalized to $c$. This relation is
identical to Eq.\thinspace 12 in
Ref\thinspace[\onlinecite{Califano1997}] for the case of two
counter-streaming, but otherwise identical electron beams. In contrast
to the growth rates predicted from the kinetic dispersion relation,
Eq.\thinspace\ref{eq:weibel:dispersion}, the cold fluid dispersion
relation predicts a \emph{larger} growth rate.

This problem requires a 3-dimensional computational domain for the
kinetic simulations to retain two velocity dimensions (1X2V). The
simulation is initialized with two homogeneous counter-streaming
populations of electrons and a neutralizing ion background (which is
not evolved during the course of simulation). The initial magnetic
field is perturbed, using a mode with wavenumber $k$. The
configuration space is periodic with a domain size of one wavelength
of the initial perturbation.

Previous work with cold fluid models\cite{Pegoraro1996, Califano1998}
suggests that the simulations should evolve into smaller and smaller
spatial scales and create magnetic field singularities. On the other
hand, the presented kinetic and two-fluid models include thermal
effects and therefore should resolve those scale lengths and saturate
when the electron gyroradius decreases to the scale of electron skin
depth.\cite{Davidson1972,Califano1998}

\begin{figure}[!htb]
  \includegraphics[width=0.9\linewidth]{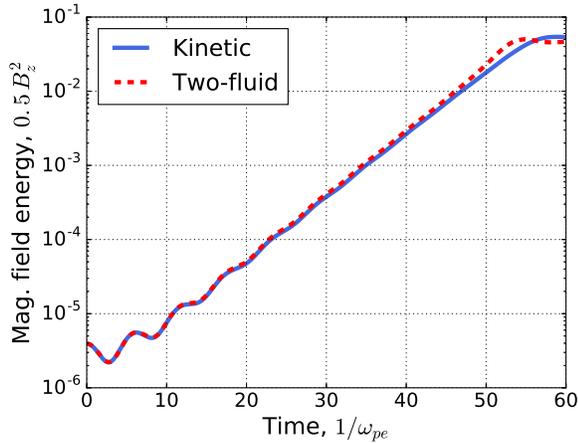}
  \caption{Growth of magnetic field energy due to the homogeneous
    Weibel instability. In this simulation two counter-streaming
    electron beams are perturbed by a small magnetic field with
    $k\lambda_D=0.04$. The fluid and kinetic models show similar
    linear growth, as well as saturate around $55/\omega_{pe}$ to
    comparable energy levels.}
  \label{fig:weibel_growth}
\end{figure}

\begin{figure}[!htb]
  \includegraphics[width=0.9\linewidth]{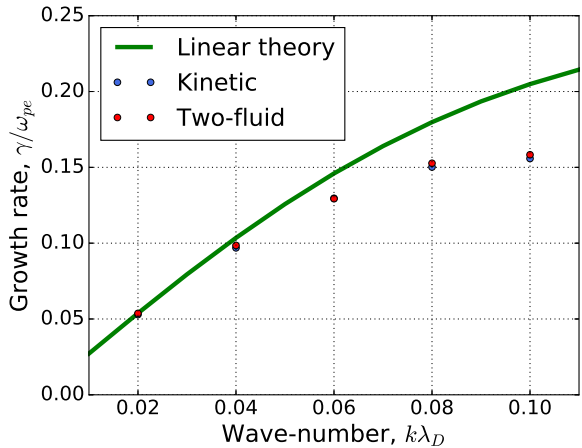}
  \caption{Comparison of the Weibel instability growth rates for the
    collisionless kinetic code, two-fluid code, and the linear theory
    prediction (Eq.\thinspace\ref{eq:weibel:dispersion}). Note that
    the initial value of thermal velocity is used in the linear theory
    dispersion relation (Eq.\thinspace\ref{eq:weibel:dispersion})
    while the simulation results show an increase in temperature
    (higher temperature corresponds to lower growth rate). Hence, the
    lack of agreement for both fluid and kinetic results for high-$k$
    modes is linked to the increase in temperature.}
  \label{fig:weibel_scan}
\end{figure}

The simulations presented in this work are run with dimensionless
parameters -- light speed $c=1$, electron temperature $T_e=0.01$,
Debye length $\lambda_{D} = 0.1$, and initial drift $v_D = \pm0.3$.
Evolution of the instability is observed in the magnetic field energy
plot in Fig.\thinspace\ref{fig:weibel_growth}.
For $k\lambda_D=0.04$ the saturation of the instability occurs around
$55/\omega_{pe}$ for both kinetic and fluid models. Growth rates,
calculated by fitting an exponential function to the energy profile,
are within 4\thinspace\% of each other and from the linear theory
prediction.  An adaptive algorithm is used for the fit by continuously
expanding the fitting region and selecting the region which produces
the fit with the highest coefficient of determination, $R^2$. This
coefficient is defined as $R^2 = 1 -
\sum_i(y_i-f_i)^2/\sum(y_i-\bar{y})^2$, where $y_i$ are data points
and $f_i$ are values of the fitted function. Results for different $k$
are in Fig.\thinspace\ref{fig:weibel_scan}.

Further study of the Weibel instability shows that plasma temperature
increases in the course of the instability growth. This increase could
explain the difference between the simulation results and the
dispersion relation (Eq.\thinspace\ref{eq:weibel:dispersion})
prediction, which uses the initial value of particle thermal
velocity. For fixed $k$ and drift velocity, the growth rate produced
by the dispersion relation decreases with increasing thermal velocity.
Detailed analysis of the Weibel instability is beyond the scope of
this work.  Further analysis of these results and of the Weibel
instability in magnetized sheaths will be presented in the future.

\section{\label{sec:setup}Problem setup and initial conditions}

It is possible to simulate a sheath in one configuration space and
only one velocity space dimension (1X1V). This setup, however, does
not capture the different evolution of parallel and perpendicular
temperatures. To accurately model the anisotropic temperature
profiles, a perpendicular temperature evolution equation
\begin{equation}\label{eq:Tperp}
	\frac{\partial}{\partial t}\left( nT_\perp\right) +
        \frac{\partial}{\partial x}\left(u_x nT_\perp\right)=\nu
        n\left(T-T_\perp\right)
\end{equation}
 needs to be solved along with the Vlasov equation. $T_\perp$ in the
 Eq.\thinspace(\ref{eq:Tperp}) stands for the temperature
 perpendicular to the flow. Eq.\thinspace(\ref{eq:Tperp}) describes
 the advection of the perpendicular temperature and its isotropization
 to parallel temperature due to collisions.

Alternatively, one may perform simulations in one configuration space
and two velocity space dimensions (1X2V) where the perpendicular
temperature is evolved self-consistently. The simulation results
presented in this paper are performed using 1X2V.  Absorbing walls are
modeled on both configuration space boundaries, and this is referred
to as a two-wall setup. While this approach doubles the simulation
cost and provides no additional information, the one-wall setup, where
one domain boundary is a free-edge and the other is an absorbing wall,
is very sensitive to the free-edge boundary condition and can lead to
unphysical results.

The configuration space ranges from -128\thinspace$\lambda_D$ to
128\thinspace$\lambda_D$, and it is divided into 256 cells. The
second-order serendipity space\cite{Arnold2011} is used for
discretization inside each cell. Second-order polynomials correspond
to three internal degrees of freedom and therefore, the Debye length
is well resolved. A realistic mass ratio of 1836 is used which
requires a different velocity space discretization for each species.
The electron velocity domain ranges from -6 to 6 initial electron
thermal speed (which is equal to $6\sigma$ of the distribution) and
ion velocity domain ranges from -5 to 5 initial Bohm speed. The
velocity space for both species is divided into 16 cells. This
resolution is chosen based on a grid convergence study using 8, 16,
32, and 64 cells, with converged results obtained for 16 cells.

For consistency with the kinetic simulations, the configuration space
of two-fluid simulations has the same range from $-128\,\lambda_D$ to
$128\,\lambda_D$ but is divided into 500 cells. The simulation is
initialized with the same density, momentum, and energy profiles as
the kinetic model. Moments of the ionization and collision terms are
included as sources in the fluid model.

The two-wall setup allows a simplified set of boundary
conditions. Dirichlet boundary conditions, $\phi=0$, are used for the
electrostatic potential at both boundaries. At the velocity space
boundaries, the particle flux is set to zero, allowing conservation of
total particle density in the domain. At the walls, the particles
streaming into the wall are completely absorbed. The wall boundary
conditions can hence be written as $f(-L/2,v_x,v_\perp,t) = 0$ for
$v_x>0$, and $f(L/2,v_x,v_\perp,t) = 0$ for $v_x<0$.

As mentioned in Sec.\thinspace\ref{sec:sheath}, simulations use a
slightly modified set of normalized variables. Even though
normalization of density to the density at the sheath edge is useful
for theoretical work, the simulation results presented here use
initial undisturbed density as the normalization.

A natural choice would be to initialize the simulation with a uniform
distribution in the configuration space and let the sheath evolve
self-consistently. This, however, leads to problems in reaching steady
state. The electric field does not appear in the presheath
instantly. As a consequence, no mechanism would accelerate the ions
from the presheath towards the wall. The ionization is, however, still
in effect and the number density in the presheath grows over the
initial value. Without density gradients or an electric field the
first moment of the Boltzmann equation (\ref{eq:Boltzmann}) simplifies
to $\partial n_s/\partial t = n_e\cdot C$, where $C$ is an constant
independent of $n_e$. This implies that when there is an imbalance,
the number density growth is exponential.  Furthermore, the initial
lack of particle flow from the presheath leads to rapid depletion of
the sheath region.

Alternatively, one may initialize the simulation with a simplified
model and let it settle into a new equilibrium with kinetic
effects. Ref\thinspace[\onlinecite{Robertson2013}] describes a model
based on the assumptions of mono-energetic ions, Boltzmann electrons,
and uniform ionization rate, $R$, over the whole domain. The validity
of the last assumption is arguable, because the ionization rate should
depend on the electron number density,\cite{Meier2012} which changes
significantly in the sheath region and is generally decreasing in the
presheath. This model is described (in dimensionless units) by the ion
momentum equation,
\begin{equation}
 \frac{\mathrm{d}\tilde{u}(x)}{\mathrm{d}\tilde{x}} = \frac{\tilde{E}(x)}{\tilde{u}(x)}-\frac{\tilde{R}}{\tilde{n}_i(x)},
\end{equation}
the Poisson equation,
\begin{equation}
  \frac{\mathrm{d}\tilde{E}(x)}{\mathrm{d}\tilde{x}} =
  \frac{\tilde{J}_i(x)}{\tilde{u}_i(x)} - e^{\tilde{\phi}(x)},
\end{equation}
and the relation between the field and the potential,
\begin{equation}
  \frac{\mathrm{d}\tilde{\phi}(x)}{\mathrm{d}\tilde{x}} =
  -\tilde{E}(x).
\end{equation}
These equations can then be integrated from the middle of the domain
towards the walls to get the density [$\tilde{n}_i =
  \tilde{R}\tilde{x}/\tilde{u}$ and $\tilde{n}_e =
  \mathrm{exp}(\tilde{\phi})$] and bulk velocity profiles. These
profiles are then used to initialize a Maxwellian distribution with
preset uniform thermal velocities.

Using these initial conditions, in the first few electron plasma
oscillation times ($1/\omega_{pe}$), excess electrons quickly leave
the domain and fluxes are equalized. This behavior results in
excitation of waves that travel through the domain, leading to an
oscillation of potential. This behavior has been noted in other
works.\cite{Lieberman2005} Spectral analysis shows that the waves are
electron waves oscillating at exactly the plasma oscillation frequency
$\omega_{pe}$. An advantage of the initial conditions mentioned above
is that starting with the approximate solution significantly limits
the excitation of these electron waves and averaging over several
plasma periods is no longer necessary.

\begin{figure}[!htb]
  \includegraphics[width=0.9\linewidth]{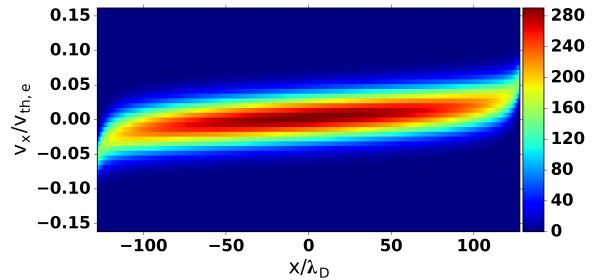}
  \caption{$xv_x$-plane of the ion distribution function evolved over
    $200/\omega_{pe}$. In this setup, there are perfectly absorbing
    walls on both boundaries of the configuration space.  This figure
    shows a noise-free solution of the ion population getting
    accelerated by the sheath electric field. Initial temperature
    ratio $T_{0e}/T_{0i} = 1$ and a realistic mass ratio $m_e/m_i =
    1/1836$ are used. The sheath potential accelerates the ions to
    supersonic speeds, adjusting to match the electron flux to give a
    net zero current.}
  \label{fig:distfIon}
\end{figure}

\begin{figure}[!htb]
  \includegraphics[width=0.9\linewidth]{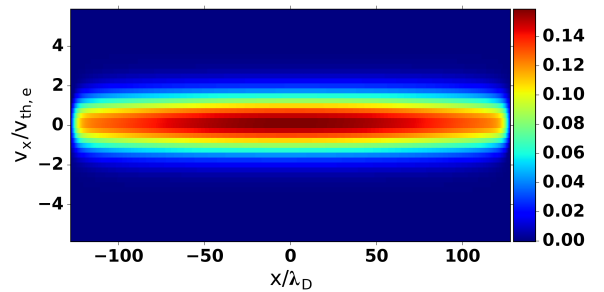}
  \caption{Same as Fig.\thinspace\ref{fig:distfIon}, expect for
    electrons. The electrons are confined by the sheath potential,
    with the high energy electrons lost to the walls.}
  \label{fig:distfElc}
\end{figure}

Phase-space plots of distributions ($xv_x$-planes of 1X2V
distributions) are presented in Fig.\thinspace\ref{fig:distfIon}
(ions) and Fig.\thinspace\ref{fig:distfElc} (electrons) using the
described model for initial conditions.\cite{Robertson2013} These
figures show smooth distribution function profiles for each of the
species.  The velocity space domain that is used is sufficient to
capture the majority of the distribution function for each of the
species including the ion distribution in the ion acceleration region.
Note some key features of sheath physics captured here -- electron
trapping near the wall and ion acceleration both in the sheath and
presheath.

\section{\label{sec:comparison}Discussion and comparison of models}

\begin{figure}[!htb]
  \includegraphics[width=0.9\linewidth]{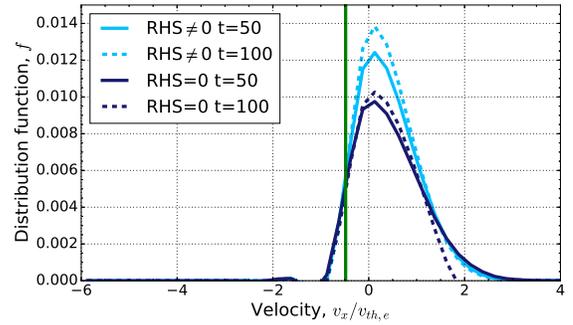}
  \caption{Electron distribution function cross-section ($v_x$) at a
    distance of $\lambda_D/3$ from the right wall. Two simulation
    results -- with collisions and ionization (RHS$\neq$0) and without
    either (RHS=0) -- are plotted here at two times, 50$/\omega_{pe}$
    and 100$/\omega_{pe}$. The velocity corresponding to the
    electrostatic potential at this location is plotted as a green
    line. Note that the distribution is in the simulation captured
    from -6\thinspace$v_{th,e}$ to 6\thinspace$v_{th,e}$.}
  \label{fig:wall_distf}
\end{figure}

As mentioned previously, continuum kinetic methods provide access to
the full distribution function everywhere in the domain.  Hence, one
can directly plot the distribution function at any point in space to
study any deviations from a Maxwellian distribution that may exist.
The distribution function cross-section for the electrons is plotted
in Fig.\thinspace\ref{fig:wall_distf} with and without the ionization
sources and collisions at two different times. Simulations marked as
RHS=0 in the figure are performed without both collisions and
ionization. All lines are for an initial temperature ratio of
$T_{0,e}/T_{0,i}=1$.  Solutions are shown for two different times,
50$/\omega_{pe}$ and 100$/\omega_{pe}$.  The distribution functions
are plotted at the right domain boundary; hence, positive velocity in
this plot denotes the outflow of electrons.  In the presence of
sources, the part of the distribution function that lies in the
positive velocity region is Maxwellian early and late in time.  The
part of the distribution function that lies in the negative velocity
region is non-Maxwellian, and the sharp gradient in the distribution
occurs due to fast electrons leaving the domain and slower electrons
being reflected by the electrostatic potential.  The vertical green
line in the plot shows a speed corresponding to the electrostatic
energy, $v = \sqrt{2 e\phi/m_e}$.  Simulations without the sources
exhibit different behavior, particularly late in time.  Without
sources, the solution (dark blue dashed line) at 50$/\omega_{pe}$
resembles the case with sources i.e. Maxwellian distribution in the
positive velocity region and abrupt drop in the distribution in the
negative velocity region.  However the later-time solution at
100$/\omega_{pe}$ (cyan dashed-line) starts to show non-Maxwellian
features in the positive velocity region of the distribution function
as well.  In the absence of collisions and ionization, the fast
electrons leave the domain, and the information propagates through the
entire domain before showing up in the distribution function on the
positive velocity region of this plot at the right boundary.  When
collisions and ionization are included, the electron distribution gets
thermalized in the bulk plasma as a result of which it remains
Maxwellian in the positive velocity region.  The negative velocity
region of the distribution function on the right wall is not affected
by the inclusion of sources (not forced to fluid-like Maxwellian)
whereas the positive velocity region is significantly altered by them.
The smoothing of the negative velocity region of the distribution
function could be caused by numerical dissipation or it could be an
inherent effect of time marching schemes. Detailed investigation of
this feature is deferred to the future work.

\begin{figure}[!htb]
  \includegraphics[width=0.9\linewidth]{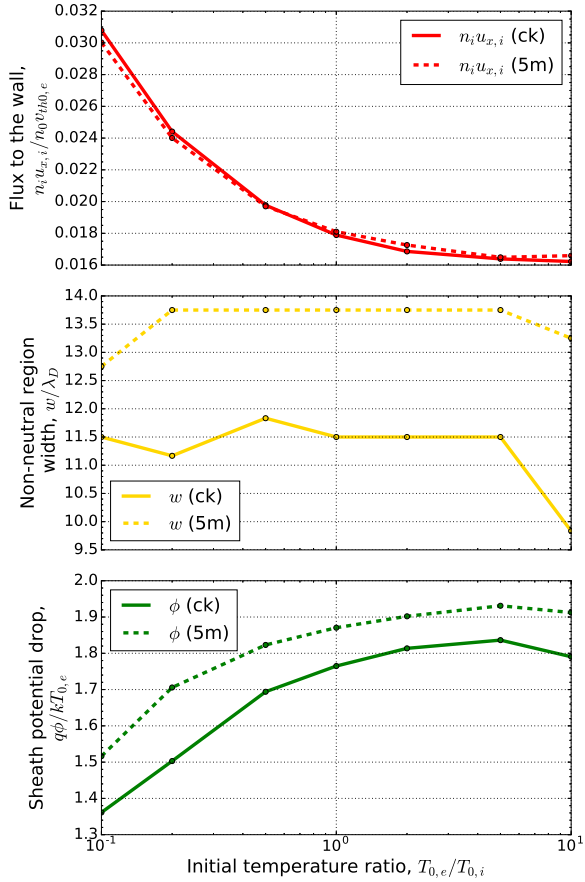}
  \caption{Comparison of the flux to the wall (\emph{Top}),
    non-neutral region width (\emph{Middle}); defined by the
    1\thinspace\% difference between electron and ion densities), and
    electric potential drop over this region (\emph{Bottom}) versus
    the initial temperature ratio between the continuum kinetic (ck)
    and five-moment two-fluid (5m) models. See
    Tab.\thinspace\ref{tab:comparison} for quantitative comparison.  Simulations are evolved from initial
    conditions for 200$/\omega_{pe}$.}
  \label{fig:scans}
\end{figure}

To explore the macroscopic effects of this kinetic behavior, the
continuum kinetic and fluid simulations are compared for several
temperature ratios. The summarizing plots of the ion momentum to the
wall, non-neutral region width, and electric
potential drop over this region evolved for 200$/\omega_{pe}$ are
presented in Fig.\thinspace\ref{fig:scans}. It is not straightforward
to the define the non-neutral region width, because the difference in
number densities of the species is introduced continuously. Instead,
1\% difference is arbitrary chosen and the distance of this point from
the wall is taken as the non-neutral region width.

Figure\thinspace\ref{fig:scans} shows that the kinetic and fluid
models provide good agreement when comparing flux to the wall at
several temperature ratios.  The distribution function is
non-Maxwellian for electrons that have been reflected by the
electrostatic potential. The part of the electron population that is
propagating towards the wall still retains the half-Maxwellian
distribution. Both models also use a realistic mass ratio and the same
temperature ratio between the species. Therefore, it is reasonable to
expect the same equalizing ion flow towards the wall after the
simulation settles into a quasi-steady-state.

The number density is defined as an integral of the distribution
function over the velocity space.  In the sheath region near the
domain boundaries, the non-Maxwellian distribution function has a
sharp gradient in the inflow part of the distribution function.  This
leads to a difference in sheath electron number density between the
kinetic and fluid models and subsequently, affects the sheath edge
calculation (marked as the non-neutral region width)
as seen in Fig.\thinspace\ref{fig:scans}.  When the plasma
distribution gets thermalized in the bulk plasma, this difference
disappears.

The electrostatic potential is compared between the fluid and kinetic
models in Fig.\thinspace\ref{fig:scans} for different temperature
ratios.  Some of this difference is attributed to the difference in
the sheath width due to which different physical locations are used to
determine the potential.  This approach is used because the potential
drop over the sheath region is a relevant physical parameter used to
characterize sheaths.

The quantitative comparison of kinetic and fluid results are
summarized in Table\thinspace\ref{tab:comparison} for the same three
parameters plotted in Fig.\thinspace\ref{fig:scans}, namely, flux to
the wall, sheath width, and electrostatic potential.  The fluid and
kinetic models agree to within about $20\%$ of each other which
provides some confidence in the boundary conditions used for the
fluids.
\begin{table}
  \centering
  \begin{tabular}{ cccc }
    \hline $T_{0e}/T_{0i}$ & Flux & Sheath width & Potential \\ \hline
    0.1 & 2.4\% & 10.9\% & 11.4\% \\ 1.0 & 1.1\% & 19.5\% & ~5.9\%
    \\ 10.0 & 2.2\% & 34.7\% & ~6.8\% \\ \hline
  \end{tabular}
  \caption{Difference between the continuum kinetic and two-fluid
    solutions (defined as $|kinetic-fluid|/kinetic$) for flux to the
    wall, quasi-neutrality region width (defined by the 1\thinspace\%
    difference between electron and ion densities), and electric
    potential drop over this region. See Fig.\thinspace\ref{fig:scans}
    for plots. Simulations are evolved from initial conditions for
    200$/\omega_{pe}$.}
  \label{tab:comparison}
\end{table}
The temperature ratios in the $x$-axis of Fig.\thinspace\ref{fig:scans}
are initial values. As particles are created by ionization at the
local temperature, the plasma cools with slightly different rates for
electrons and ions since there is no source of heating in these
simulations.

\begin{figure}[!htb]
  \includegraphics[width=0.9\linewidth]{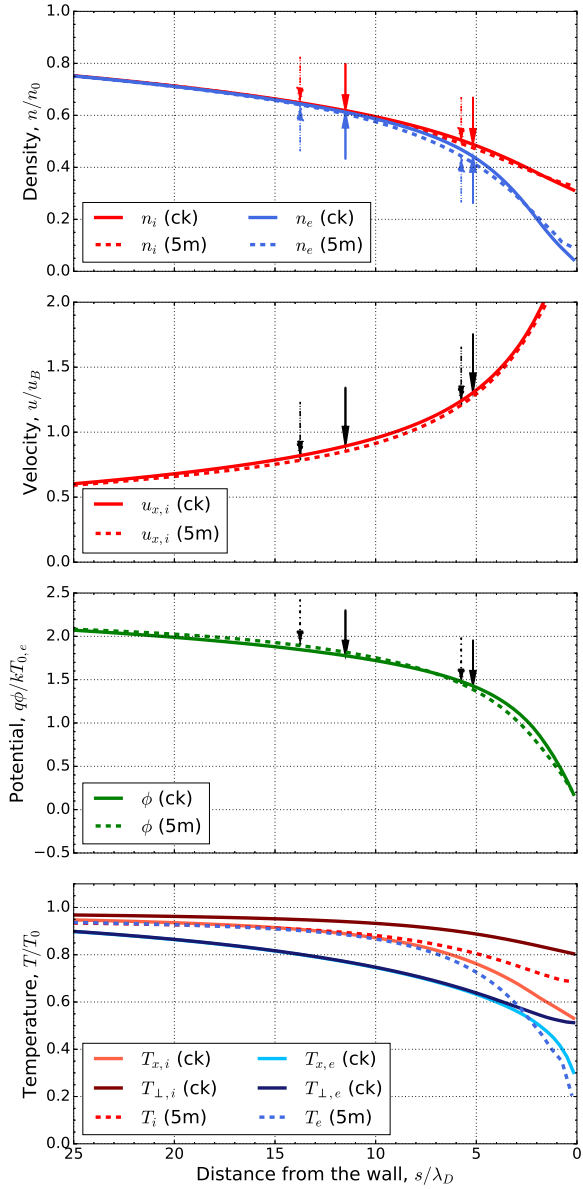}
  \caption{Comparison of continuum kinetic (ck) and five-moment
    two-fluid (5m) simulations for initial $T_{0e}/T_{0i} =
    1$. \emph{Top} Densities are normalized to the initial density in
    the middle of the domain. \emph{Upper middle} Velocities
    normalized to the local modified Bohm velocity ($u_B =
    \sqrt{\mathrm{k}_\mathrm{B}\left(T_{x,e}+T_{x,i}\right)/m_i}$). \emph{Lower
      middle} Electrostatic potentials normalized to
    $\mathrm{k}_\mathrm{B}T_e$.  \emph{Bottom} Parallel ($T_x$) and
    perpendicular temperatures. Results of the kinetic code are
    plotted in the solid line, the two-fluid code in the dashed
    line. Electron quantities are plotted in blue and ion in
    red. Arrows show points where a relative difference between
    electron and ion densities are bigger than 1\thinspace\% and
    10\thinspace\% (solid arrows correspond to kinetic code, dashed to
    fluid code). Simulations are evolved from initial conditions for
    200$/\omega_{pe}$.}
  \label{fig:comp}
\end{figure}

Detailed profiles of normalized density, velocity, electrostatic
potential, and temperatures near the wall are presented in
Fig.\thinspace\ref{fig:comp} for $T_{0e}/T_{0i} = 1$ and after
200/$\omega_{pe}$. The densities for both species are normalized to
the initial density, velocities to the local modified Bohm velocity,
$u_B = \sqrt{\mathrm{k}_\mathrm{B}\left(T_{x,e}+T_{x,i}\right)/m_i}$,
and the electrostatic potential is normalized to
$\mathrm{k}_\mathrm{B}T_{0e}$. Locations where the plasma is no longer
quasi-neutral (1\thinspace\% and 10\thinspace\% difference in electron
and ion densities) are marked by arrows, with solid arrows
representing locations for the kinetic model and dashed arrows
representing locations for the fluid model.  It is worth noting here
that the crossing of the Bohm speed corresponds to the point with 1\%
difference in number density.  The definition of the sheath width used
is, therefore, consistent with classical sheath theory.

\begin{figure}[!htb]
  \includegraphics[width=0.9\linewidth]{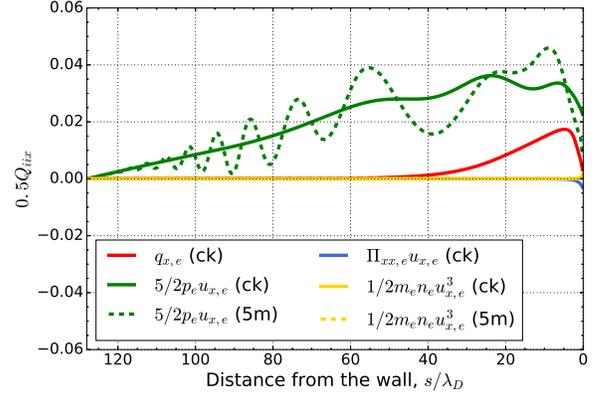}
  \caption{Individual terms of the expanded heat flux
    (Eq.\thinspace\ref{eq:Q}). Heat flux vector in the plasma frame,
    $q$ and stress tensor are only available in the continuum kinetic
    simulations (marked as ck). Simulations are evolved from initial
    conditions for 100$/\omega_{pe}$.}
  \label{fig:Q}
\end{figure}

The temperatures plotted in the bottom subfigure of
  Fig.\thinspace\ref{fig:comp} provide an interesting comparison
  between fluid and kinetic results and highlight the importance of
  kinetic effects.  The five-moment two-fluid model used here assumes
  an isotropic temperature. The parallel temperatures ($T_x$) for
both electrons and ions in the continuum kinetic results undergo
decompression cooling as expected.\cite{Tang2011} The parallel
temperature is then equalized with the perpendicular temperature
through collisions.  It is observed that this effect is more apparent
for electrons due to their higher collision frequency with respect to
ions.  The ion temperature is in good agreement between continuum
kinetic and fluid simulations (isotropic fluid temperature lies in
between the ion parallel and perpendicular). However, the electron
temperature has more significant differences between the kinetic and
fluid results.  To understand this, higher moments are needed.  The
second moment of the Vlasov equation leads to the energy conservation
equation
\begin{equation}
  \frac{\partial \mathcal{E}}{\partial t} + \frac{1}{2}\frac{\partial
    \mathcal{Q}_{iik}}{\partial x_k} = nq\textbf{u}\cdot\textbf{E},
\end{equation}
where 
\begin{align}
  \mathcal{E} = \frac{3}{2}p + \frac{1}{2} mn\mvec{u}^2
\end{align}
is the particle energy and
\begin{align}
  \mathcal{Q}_{ijk} = m \int v_i v_j v_k f \thinspace d^3\mvec{v}.
\end{align}
is the heat flux tensor. Contraction of $\mathcal{Q}_{ijk}$ gives
(twice) the particle energy-flux density and can be expanded as
follows
\begin{equation}\label{eq:Q}
  \frac{1}{2}\mathcal{Q}_{iix} = \underbrace{q_x + u_x\Pi_{xx}}_{non-ideal} +
  \underbrace{\frac{5}{2}pu_x+\frac{1}{2}mnu_x^3}_{ideal},
\end{equation}
where $\Pi_{xx}$ is the parallel component of the stress tensor, $p$
is pressure, and
\begin{equation}
  q_x =
  \frac{1}{2}m\int_{-\infty}^\infty\int_{0}^\infty\left(w_x^2 +
  v_\perp^2\right)w_xf(v_x,v_\perp)2\pi v_\perp dv_\perp dv_x
\end{equation}
is the heat flux vector in the plasma frame, and $w_x =
v_x-u_x$.
Individual terms of Eq.\thinspace(\ref{eq:Q}) are plotted in
Fig.\thinspace\ref{fig:Q} for the electrons. The
  decompression cooling terms (green lines) between the kinetic and
  fluid results are in good agreement and are dominant terms.  The
  five-moment fluid model used here does not capture the kinetic
  physics of the heat flux vector and the stress tensor.  The stress
  tensor plotted in Fig.\thinspace\ref{fig:Q} (blue line close to
  zero) is negligible in this case and is not a significant kinetic
  effect here.  The heat flux vector (red line), however, is
  significant in the sheath.  The heat flux vector is negligible in
  the bulk plasma where the distribution function is thermalized by
  collisions.  Within about $50$ Debye lengths of the wall, however,
  the heat flux becomes significant and this could explain the
  differences in electron temperature between the kinetic and fluid
  results.

\section{\label{sec:conclusions}Conclusions}

The continuum kinetic solvers using the discontinuous Galerkin scheme
in the \gke\ code might provide access to high-dimensionality
simulations and noise-free results which have been inaccessible thus
far. Successful benchmarks of the scheme have been performed and
compared to previous literature as well as to fluid simulations.

The kinetic and fluid models are benchmarked with the collisionless
Weibel instability and good agreement is found.  Additionally, the
saturation of the fluid Weibel instability is converged and
corresponds to the kinetic saturation. The extracted growth rates
agree with each other and with the linear theory prediction for small
wavenumbers. The mechanism for nonlinear saturation and effect of
increasing temperature on linear growth rates will be further explored
in future work.

A 1D classical sheath using a two-wall setup is simulated using both
the kinetic and two-fluid scheme. The kinetic and two-fluid results
are in good agreement for momentum and density of both species over
the sheath region (see Table\thinspace\ref{tab:comparison} for
quantitative comparison), demonstrating that the two-fluid model may
be useful in certain regimes to study sheath physics.  
  However, key differences are present when looking at higher moments
  such as parallel ($T_x$) and perpendicular temperatures where the
  5-moment fluid model assumes an isotropic temperature. These
  differences in the temperature are attributed to the heat flux
  vector, a kinetic effect which is missed in the fluid model. 
Additionally, the distribution function in the sheath is
non-Maxwellian highlighting the need for kinetic physics.  The
inclusion of further physics and, in particular, magnetic fields
oriented at arbitrary angles to the wall may lead to further
differences between the fluids and kinetic models due to finite orbit
effects, not typically captured completely in the fluid model.

The models described here will be extended to include surface and
atomic physics, with more sophisticated collision terms, which are
sensitive to the non-Maxwellian shape of the plasma distribution
function near the wall, thus requiring a kinetic model. Also, the
inclusion of magnetic fields in the kinetic model will provide a tool
to study plasma/solid-surface interactions relevant to Hall thrusters,
as well as develop parameterized boundary conditions, along the lines
of Ref\thinspace[\onlinecite{Loizu2012}], for use in fluid
simulations of fusion machines.

\begin{acknowledgments}
  This research was partly supported by the Air Force Office of
  Scientific Research under grant number FA9550-15-1-0193. The work of
  A.H. was supported by the U.S. Department of Energy through the
  Max-Planck/Princeton Center for Plasma Physics, the SciDAC Center
  for the Study of Plasma Microturbulence, and Laboratory Directed
  Research and Development funding, at the Princeton Plasma Physics
  Laboratory under Contract No. DE-AC02-09CH11466.

  Useful discussions with Greg Hammett of PPPL and Wayne Scales of VT
  are acknowledged.
\end{acknowledgments}

\bibliography{reference}

\end{document}